\documentclass{aa}  
\usepackage{graphicx}
\usepackage[colorlinks,allcolors=blue]{hyperref}
\bibpunct{(}{)}{;}{a}{}{,}
\usepackage[varg]{txfonts}
\usepackage{xcolor}
\usepackage{mathtools}
\usepackage{subfig}
\usepackage{graphicx}
\usepackage{footmisc}
\usepackage{amssymb}
\usepackage{amsmath}
\usepackage{graphicx}
\usepackage{booktabs}
\usepackage{amsfonts}     
\usepackage[T1]{fontenc} 
\usepackage[utf8]{inputenc} 
\usepackage{multicol}
\usepackage{xcolor}
\usepackage{soul}
\usepackage{rotating}
\usepackage{comment}
\usepackage{multirow}
\usepackage{longtable}
\usepackage{lscape}
\usepackage{color}
\usepackage{url}
\usepackage{url}

\usepackage[normalem]{ulem}

\def\apjl{\rm ApJL}
\def\apjs{\rm ApJS}

\def\aap{\rm A\&A}

\begin{document}

   \title{
   Secondary black hole-induced magnetic reconnection in OJ 287: Implications for X-ray and radio emission}


   \author{S. Boula
          \inst{1}\orcid{0000-0001-7905-6928}
          \and
         A. Nathanail \inst{2}\fnmsep\thanks{corresponding author}\orcid{0000-0002-1655-9912}
          }

   \institute{INAF – Osservatorio Astronomico di Brera, Via E. Bianchi 46, I-23807 Merate, Italy\\
              \email{styliani.boula@inaf.it}
             \and 
                          Research Center for Astronomy, Academy of Athens, GR-11527, Athens, Greece\\
             \email{anathanail@academyofathens.gr}
             }

   \date{Received XXX XX, XXXX; accepted XXX XX, XXXX}

 
  \abstract
   {OJ 287, a nearby blazar, has exhibited remarkable variability in its optical light curve since 1888, characterized by $\sim12$-year quasi-periodic outbursts. These events are attributed to the orbital dynamics of a supermassive binary black hole system at the heart of the blazar. This study explores the role of magnetic reconnection and the formation of plasmoid chains in driving the energetic processes responsible for OJ 287's variability. We propose that the passage of the secondary black hole through the magnetic field of the primary black hole's accretion disk triggers magnetic reconnection, which contributes to the observed X-ray and radio emission features in OJ 287.%
   }
   {We explore the connection between binary black hole interactions, accretion disk dynamics, and the formation of plasmoid chains as the secondary black hole passes through the magnetic field forest from the accretion disk and the jet of the primary. Magnetic reconnection is the fundamental process behind particle acceleration, potentially influencing the observed emissions and variability, particularly during specific orbital phases of OJ 287.}
   {Our approach relies on numerical simulations to understand the formation of plasmoid chains resulting from black hole interactions and accretion disk dynamics. Based on such results, we employ simulation outcomes to examine the potential contribution to observed emissions, validating our assumptions about plasmoid chain creation. With this idea, we aim to establish a direct link between numerical simulations and observed emission, particularly in the case of OJ 287.}
   {Our findings confirm that the formation of plasmoid chains coincides with specific anomalous emission events observed in OJ 287. Notably, the radio emission patterns cannot be explained by a single blob model, as the necessary size to mitigate synchrotron self-absorption would be too large. This highlights the complexity of the emission processes and suggests that plasmoid chains could contribute to additional emission components beyond the steady jet}.
   {}

   \titlerunning{Magnetic Reconnection in OJ 287}
 
   \authorrunning{Boula et al.}
   \maketitle
%
\section{Introduction}
Blazars, particularly flat spectrum radio quasars and BL Lac objects, 
are a unique class of active galactic nucleus (AGN) characterized by 
relativistic jets that point directly toward Earth. The relativistic motion 
amplifies their emission across the electromagnetic spectrum, making them 
some of the most luminous and distant sources in the Universe 
\citep{Blandford2019}. Their spectral energy 
distribution (SED) exhibits two distinct “humps:” one extending from radio to 
optical-UV (and sometimes into the X-ray) bands, and another from X-rays to 
gamma rays. The lower-energy component is typically attributed to synchrotron 
emission from nonthermal electrons, while the higher-energy hump is often 
explained by inverse Compton scattering of either synchrotron photons or 
external seed photons.

OJ 287, a well-studied blazar at a redshift of $z = 0.306$,  exhibits quasiperiodic optical flares with a remarkable $\approx 12$-year periodicity, attributed to 
its binary black hole nature \cite{Sillanpaa1988}. The system's primary black hole has a mass of \( 1.8 \times 10^{10} M_{\odot} \), while the secondary has a mass of \( 1.5 \times 10^8 M_{\odot} \).  However, \cite{Komossa2023} put forward a different estimate based on a new theory of energy production in accretion disks. In this theory, all the accretion power is converted into optical, thermal emission from the accretion disk, resulting in luminosities about 100 times higher than the ones predicted by standard models. To reconcile this discrepancy, the authors propose reducing the black hole mass by a factor of 100.  Reducing the black hole mass by a factor of 100 was suggested to address this inconsistency. Notably, this discrepancy is not unique to OJ 287 but extends to other AGNs as well, as is discussed by \cite{2023Valtonenb}. The orbit is eccentric and tilted, with a semimajor axis of \( 9300~\text{AU} \) \citep{Laine2020}. 
The current model, in 
which a secondary supermassive black hole (SMBH) periodically plunges through 
the accretion disk of the primary SMBH, provides an explanation for the 
optical variability observed over more than a century {\citep{Valtonen2008,2016Valtonen}}. {In the current theory by \citep{1996Lehto,1998Ivanov,Valtonen2019}, the role of the magnetic fields is neglected. However, in the recent work by \cite{Valtonen2023}, the need for the magnetic field 
 is introduced as an early stage of the bubble escaping from the perturbed accretion disk. Furthermore, in the pre-flare activity of OJ287, as was reported by \cite{Pihajoki2013}, the magnetic fields could play a role. }

To address this, we propose that magnetic reconnection, triggered by disrupting the 
accretion disk's magnetic field during the secondary's passage, plays a key role in producing the 
nonthermal radiation observed at higher energies. As the secondary black hole disturbs the disk and moves close to the jet of the primary, 
the interaction between the magnetic fields leads to the formation of current sheets and plasmoids --
magnetic islands that evolve and release energy through reconnection. These plasmoids, in turn, 
accelerate particles to relativistic energies. The particles can then radiate synchrotron and inverse 
Compton emission from the radio to X-ray regimes \citep{Guo2014, Sironi2014, BAO2018}.

Observationally, the emission mechanism responsible for certain deviations in the V/X flux 
ratio of OJ 287 remains unclear. The V/X method \citep{Valtonen2023} has identified anomalous flux 
ratios at specific orbital phases, suggesting an additional, non-jet-related emission component. In 
this work, we investigate whether magnetic reconnection, induced by the secondary’s passage, could 
be responsible for this variability. This hypothesis is tested through comparisons with observed 
multiwavelength data, which we discuss in Section \ref{sec:results}

Our model focuses on the evolution of these plasmoids and plasmoid chains\footnote{One can produce and energize a whole series of plasmoids along
the current sheet, i.e., a “plasmoid chain.”}, which are well suited for 
explaining the nonthermal flares observed in OJ 287. The key is understanding how the secondary black hole's passage through the primary's magnetized accretion disk disrupts the magnetic 
field, generating localized regions of enhanced magnetic energy. In these regions, high 
magnetization ($\sigma = B^2/\rho c^2$) and turbulent conditions lead to efficient particle 
acceleration, producing relativistic electron distributions capable of emitting synchrotron 
radiation \citep{Petropoulou2016, Meringolo2023}. 
For OJ287, this process mirrors the behavior of AGN jets, in which 
magnetic reconnection has been shown to be a critical factor in shaping the SED and variability of 
emissions \citep{Gomez2022, Valtonen2024}.
Recent particle-in-cell simulations provide a framework for quantifying the energy release 
and particle acceleration mechanisms involved in magnetic reconnection events. These simulations 
demonstrate that plasmoids form naturally in turbulent, high-magnetization environments and can 
accelerate particles to ultra-relativistic speeds \citep{Sironi2016, 
Werner2018, LGLL2023}. The rate of magnetic reconnection and the size of the plasmoids are directly 
linked to the local magnetic field strength, which in turn scales with the properties of the black 
hole system, such as the mass and spin of the primary SMBH, and the cause of local turbulence 
that is the secondary SMBH. As energetic particles trapped within the plasmoid chains gain high energies, interactions with the surrounding environment cause energy loss, emitting $\gamma$-ray, X-ray, and near-infrared flares \citep{Valtonen2023}.

In addition, general relativistic magnetohydrodynamic (GRMHD) simulations at increasingly higher 
resolutions have begun to capture the formation of current sheets and magnetic reconnection near
black holes \citep{Ripperda2020, 2020NFPOYMR, Chatterjee2021}. These 
studies highlight how magnetic dissipation in the turbulent disk environment leads to rapid 
reconnection, which is essential for understanding the variability and energetics of blazar flares. 
The insights gained from these simulations are crucial for our model, as they provide a detailed 
picture of how plasmoids evolve and interact with the surrounding plasma, ultimately driving the 
observed nonthermal emissions.

This paper is organized as follows. In Section \ref{sec:model}, we
sketch the proposed model, with subsection \ref{sec:plasm}
focusing on the development of plasmoids and subsection \ref{sec:emis}
reporting the details of the emission process. In section \ref{sec:results},
we analyze and discuss the results of our calculations, 
and in Section \ref{sec:con} we present our conclusions.

\section{Plasmoid driven variability in OJ 287  -  \newline model details}
\label{sec:model}
\begin{figure}
    \centering    
 
    \includegraphics[width=0.94\linewidth]{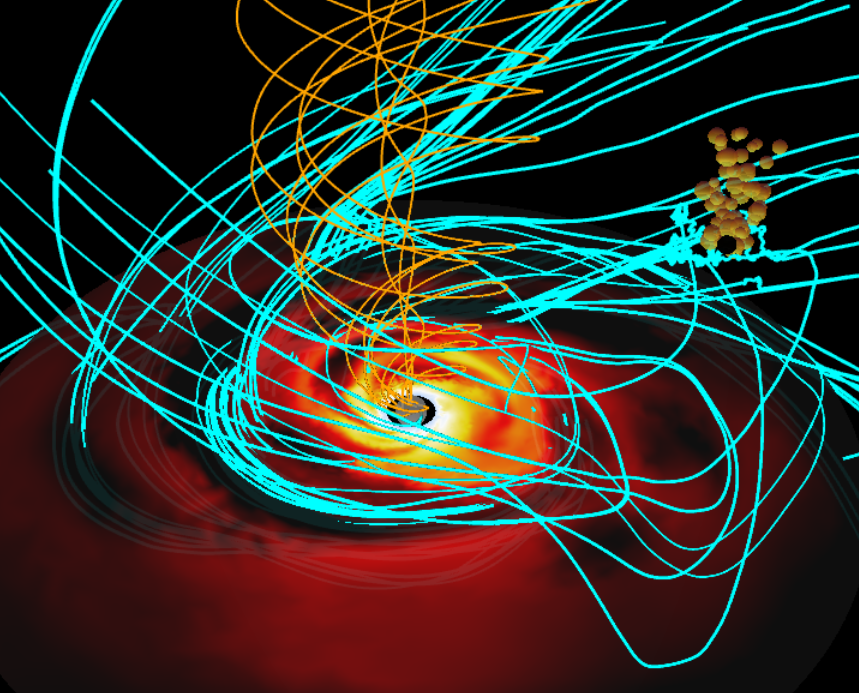}

    \caption{3D cartoon representation of the accretion 
    flow and magnetic field topology around the primary black 
    hole. The rest-mass density distribution is shown in the 
    equatorial plane using a yellow-to-red color gradient. The 
    jet of the primary black hole is represented by the orange 
    magnetic field lines, while the disk’s magnetic field lines 
    are depicted in cyan, highlighting the regions where 
    magnetic reconnection is expected to occur, leading to 
    plasmoid formation above the accretion disk. The secondary 
    black hole is illustrated as a black circle above the 
    equatorial plane, with the gold blobs representing the 
    expected plasmoids generated by the perturbation of the 
    disk's magnetic field due to the secondary black hole’s 
    passage.
Note that the primary black hole, the disk, and the magnetic 
field lines were derived from a 3D simulation \citep{2022NMPFR}, 
with the secondary black hole and plasmoids 
added ad hoc.
}
    \label{Fig:mhd}
\end{figure}

The accretion disk surrounding the primary black hole in OJ 287 is expected 
to be magnetized. First, the primary black hole is known to possess a 
relativistic jet with a helical magnetic field \citep{Gomez2022}, 
suggesting that the magnetic field at the horizon must have been accreted 
through the inner regions of the disk. This implies that the accretion disk is 
likely threaded by a large-scale poloidal magnetic field that extends to 
high altitudes closer to the jet,
where the density is expected to be low. Second, this large-scale magnetization is supported by observations and models of jet 
acceleration and collimation zones, where magnetic fields play a crucial role 
\citep{Gold2014, Mizuno2014}. Moreover, previous studies have explored the 
complex dynamics of the binary system and the role of magnetic fields in 
triggering various types of emissions. \citealt{Pihajoki2013} specifically 
highlighted how the magnetic interaction between the black holes could drive 
emissions across different wavelengths. Additionally, polarization data 
from OJ 287 indicate that the inner jet follows a helical trajectory, 
consistent with the presence of a helical magnetic field component within the 
jet's core, as is seen in VLBI (Very-long-baseline interferometry) observations\citep{Myserlis2018}.

Building on this, our model proposes that, although the magnetic field in the 
accretion disk of the primary black hole is likely less intense than the jet’s 
field, it still retains a helical structure. As the secondary black hole 
approaches and passes through the disk, 
it disturbs the disk’s magnetic geometry and the magnetic field close to the jet, 
inducing shocks 
and triggering intense magnetic reconnection. This reconnection process leads 
to the formation of plasmoids and plasmoid chains. From the perspective of the 
secondary black hole, it would appear as though an enormous “magnetic forest” 
were being accreted from one direction, in contrast to the typical inward 
rotation of matter in accretion disks. Close to the event horizon of the 
secondary black hole, regions of high magnetization ($\sigma>> 1$) would 
develop, energizing particles and trapping them within plasmoids. 

The variability introduced by this reconnection-driven process is further supported by the anomalous trends observed in the V/X method analysis \citep{Valtonen2023}, which deviates from the expected jet emission ratio. These results, discussed in Section \ref{sec:results}, suggest that plasmoid formation and evolution contribute to the observed X-ray and optical variability during the secondary’s passage through the disk and near the jet.

{This scenario is based on two high-resolution simulations: one modeling the accretion disk of the primary black hole and the other simulating plasmoid production near a black hole. }
{The dynamics described here can be better understood by visualizing recent simulations }(see Fig. \ref{Fig:mhd}
\footnote{The magnetic field around the black 
a random vector field has perturbed hole with a 
Gaussian profile to model the influence of the secondary black 
hole on the disk’s magnetic structure.}), {where the secondary black hole can be seen entering this complex magnetic environment. Details on the numerical simulations can be found in Appendix A.}


\subsection{Plasmoid chains}
\label{sec:plasm}

Plasmoids form in highly magnetized environments, serving as efficient sites for particle acceleration, which produces high-energy emissions consistent with the variability observed in blazar jets \citep{Sironi2016}. This process is central to our model for OJ 287, where the passage of the secondary black hole through the primary’s magnetized accretion disk induces magnetic reconnection, triggering plasmoid formation. These plasmoids can explain the nonthermal X-ray and radio flares observed in the system, with plasmoid mergers generating multiwavelength flares and polarization variations \citep{Zhang2024}.

For the primary black hole in OJ 287, the gravitational radius and timescale are given by
$r_{\rm g1} = \frac{GM_1}{c^2} = 10^{15.26} \, \text{cm}, \quad t_{\rm g1} = \frac{r_{\rm g1}}{c} = 10^{4.78} \, \text{s},$
with the Eddington luminosity,
$L_{\rm Edd} \approx 2.27 \times 10^{48} \left(\frac{M}{M_1}\right) \, \text{erg} \, \text{s}^{-1}.$
For the secondary black hole, the relevant timescale is $t_{\rm g2} = 10^{2.5} \, \text{s}$.
Recent simulations show plasmoid formation at the boundaries of magnetized funnel regions, with plasmoids typically produced every \(500 - 700 \, t_{\rm g}\) \citep{2020NFPOYMR, Ripperda2020, Ripperda2022, 2022NMPFR}. The plasmoid production rate can be approximated by:
\begin{equation}\label{dotN}
\dot{N}_{\rm plasmoid}\approx \frac{10^3}{2\times 10^5\, t_{\rm g2}}=1/500\, t_{\rm g2}^{-1}
,\end{equation}
with larger plasmoids forming every 500 \( t_{\rm g2} \) and smaller plasmoids occurring more frequently. Each one of these plasmoids contributes significantly to the overall emission. Plasmoid production is derived from a high-resolution simulation that studies  how matter plunging into a black hole leads to plasmoid generation due tothe  bringing of inverse magnetic field polarity together and forming turbulence \citep{2020NFPOYMR}. 
This study informs our assumptions about plasmoids' energetics, morphology, and evolution during the secondary black hole’s passage through the magnetized environment of the primary’s accretion disk.

A single flare event in  OJ 287 has been observed to release between $10^{47}$ and 
$10^{48}$ ergs, depending on the wavelength \citep{Marscher2011, Valtonen2019}. 
The minimum energy budget for each plasmoid can be 
estimated by dividing the total flare energy by the number of plasmoids, giving a rough 
estimate of $10^{44} - 10^{45}$ ergs per plasmoid. While plasmoids possess more magnetic 
energy than this, only a fraction is radiated in the observed frequencies. 
The magnetic energy of each plasmoid can be estimated as
\begin{equation}\label{EB}
E_{\rm B, plamoid}\approx 3.3 \times 10^{46} \left( \frac{r}{2\,\,r_g} \right)  
\left(\frac{B}{2\,{\rm G}}\right) \, \, {\rm  erg}.
\end{equation}
These estimates support the 
idea that a thousand plasmoids can collectively produce the required energy for a flare. More details on the production of plasmoid chains in numerical simulations of black hole accretion can be found in Appendix \ref{appB}. The properties of the plasmoid chains, including their formation timescales, sizes, and expansion velocities, are closely related to the mass and size of the black hole.
The energy distribution of electrons within the plasmoids follows a power law,  characterized by a high-energy tail contributing to the nonthermal radiation observed in X-rays and radio. Understanding these distributions and scalings  with the black hole parameters provides a comprehensive model for the energy budget of OJ 287’s flares. 
While nonthermal radiation from dual jet interactions between two SMBHs has been discussed previously \citep{Gutierrez2024}, we focused on the interaction between the secondary black hole and the primary disk’s magnetic field, as this interaction can produce the observed flares.

{To investigate the role of plasmoid production in OJ 287, we considered the interaction of the secondary black hole with the primary’s accretion disk. The matter distribution of the primary’s disk, extracted from a high-resolution simulation, provides a reliable basis for our analysis. This simulation accurately captures the density profile and turbulence within the disk, forming the foundation for our understanding of the secondary’s interaction with the disk material. The simulation details can be found in the Appendix \ref{appA}. As was mentioned earlier, the passage 
of the secondary black hole through the accretion disk of the primary is similar to the procedure of matter plunging onto a stationary black hole, if we move to the rest frame of the secondary black hole, and this situation has been simulated and the results used here \citep{2020NFPOYMR}.}

Figure \ref{Fig:rho} shows 1D slices of the logarithmic density, $\log_{10}(\rho) $, along the height, $z $, of the accretion disk at various radial distances, $ x $, from the SMBH, with $y = 0 $. Each curve corresponds to a specific value of $ x $ (20, 30, 40, 50, and 60). The density profile is maximal at $z = 0 $, corresponding to the disk midplane. However, additional density peaks are observed at higher $z $ values, indicating vertical density variations that may be linked to disk turbulence or structural instabilities. These peaks diminish in magnitude as $ x $ increases, suggesting a radial dependency of density stratification within the disk. 
{The height, $ z $, can  be understood in terms of the secondary black hole’s trajectory through the primary’s disk.}  Specifically, phase 1 begins when the black hole enters the disk and then passes through the maximum density region at $z = 0 $, coinciding with the X-ray flare observation. Phase 2 occurs as the black hole exits the interaction region.

\subsection{Emission}
\label{sec:emis}
In the following, we introduce a model based on a chain of plasmoids, which may contribute to particle acceleration and the generation of nonthermal populations. We used simple assumptions as a zero-order approximation to determine whether this model can account for the observed emission data.
 The broader morphology of the nonthermal SED depends
on the ratio of the magnetic to photon energy densities. To calculate the former, we assumed some equipartition with the accreting matter, the energy density. By assuming the power of the accretion, $P_{\rm acc}= \dot{m}{ M}L_{\rm Edd}$, with  $\dot{m}$ the mass accretion rate normalized to the Eddington one and 
${ M}= M_{1}/M_{\rm \odot}$ (recall that
$M_1$ is the mass of the primary black hole).
We assume that $\dot{m}\approx 0.1$, which is consistent 
with other studies for OJ 287 \citep{Marscher2011, Valtonen2019}.
For example, the nonthermal emissivity, which arises from the synchrotron radiation of relativistic electrons, can be expressed, as in \cite{RL1979}, as
\begin{equation}
    j_{\nu} \propto \sigma_T c K U_B \nu_L \left( \frac{\nu}{\nu_L} \right)^{-\frac{p+1}{2}}
,\end{equation}
where $\sigma_T$ is the Thomson cross-section, $c$ is the speed of light, $K$ is the normalization factor of the electron distribution that is connected to the electron density, $U_B$ is the magnetic energy density, and $\nu_L$ is the Larmor frequency, given by  $\nu_L = \frac{e B}{2\pi m_e c}$. According to \cite{Ball2018}, the slope of the spectrum concerning the magnetization, $\sigma$, and plasma, $\beta$, is given by
   $ p = A_p + B_p \tanh (C_p \beta)$,
where  
$    A_p = 1.8 + \frac{0.7}{\sigma}, \quad B_p = 3.7\sigma^{-0.19}, \quad C_p = 23.4\sigma^{0.26}$. Moreover, the electron nonthermal efficiency with respect to the magnetization, $\sigma$, and plasma, $\beta$, is  $ \epsilon = A_{\epsilon} + B_{\epsilon} \tanh (C_{\epsilon} \beta)$,
where  $    A_{\epsilon} = 1 - \frac{1}{4.2\sigma^{0.55} + 1}, \quad B_{\epsilon} = 0.64\sigma^{0.07}, \quad C_{\epsilon} = -68\sigma^{0.13}$.
To model particle radiation, we calculated the electron distribution fraction from the kinetic equation, 
which describes the losses and/or sinks and injection of relativistic electrons from the acceleration zone ($Q_e$) \cite{BM2022}, more details in the Appendix \ref{appC}. 
\begin{table}[h!]
\centering\caption{Average parameter values of plasmoids.}
\begin{tabular}{|c|c|}
\hline
Parameter & Value \\ \hline\hline
Initial radius $R_{in}~(r_{\rm g})$ & 3.5  \\ \hline
Initial Magnetic field $B_{in}~({\rm G})$ & 2 \\ \hline
Expansion velocity $u_{exp}~(c)$& 0.03 \\ \hline
Bulk Lorenzt factor $\Gamma$& 5 \\ \hline
Duration $\Delta t~(t_{\rm g})$& 500-1000 \\ \hline
Averaged number of plasmoids $\#$& 2000  \\ \hline\hline
 
 Electron luminosity $L_{e,in}~{\rm erg/sec}$ per plasmoid& $10^{42.8}$  \\ \hline
Minimum electron Lorentz factor $\gamma_{\rm min}$& 1 \\ \hline
Maximum electron Lorentz factor $\gamma_{\rm max}$ & $10^{4.5}$  \\ \hline
 Slope of the electron energy distribution $s$ & 2.7 \\ \hline
\end{tabular}
\label{tab:1}
\end{table}

\begin{figure}[t]
    \centering    \includegraphics[width=1\linewidth]{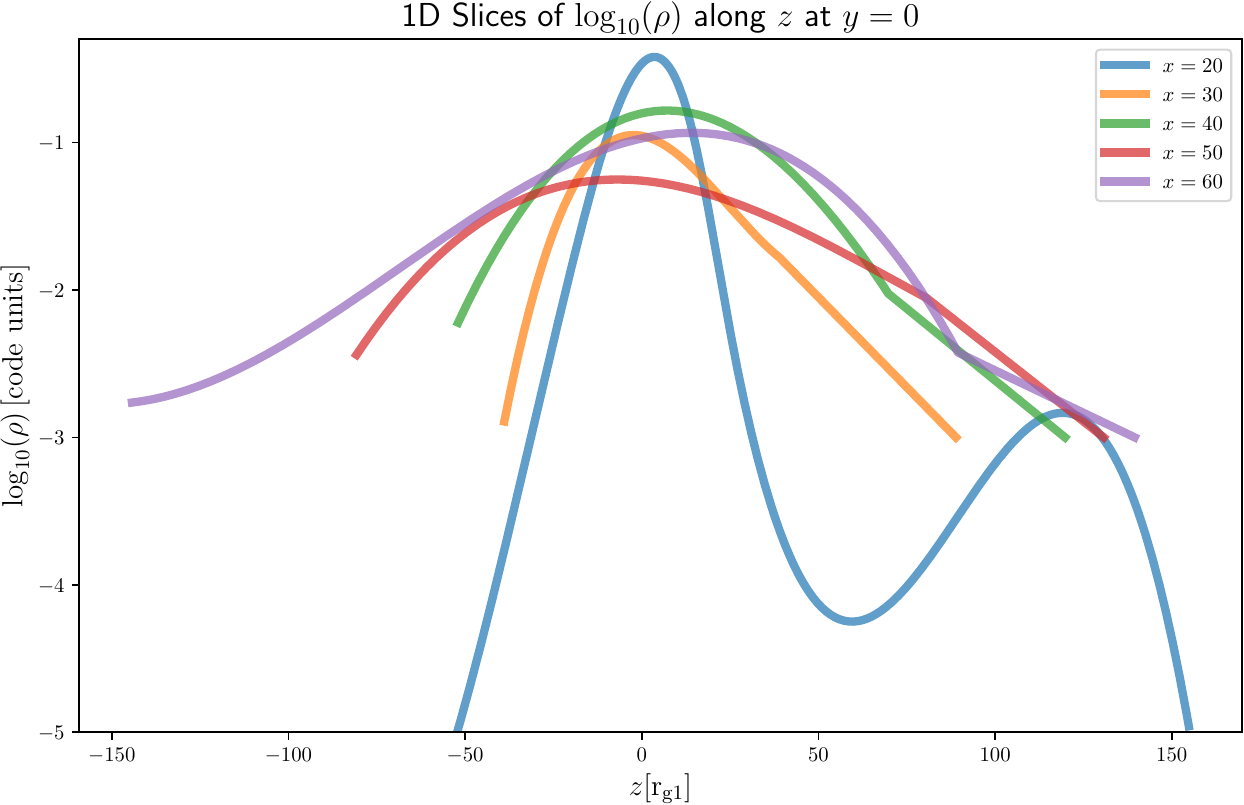}
    \caption{  1D slices of the logarithmic density, $ \log_{10}(\rho) $, along the vertical height, $z $, of the accretion disk at various radial distances, $ x $, from the SMBH, with $y = 0 $. The density peaks at the disk midplane, $ z = 0 $, but additional maxima at higher or lower heights reveal the vertical structure of the disk.}
    \label{Fig:rho}
\end{figure}

Observational evidence reveals that the X-ray emission before and after the flare remains almost steady, suggesting that it originates from the accretion disk \cite{Komossa2021}. In contrast, the variability observed in the radio emission points to the presence of a nonthermal particle distribution. We employed a simplified approach -- although stochastic models can also be used -- to reproduce this radio variability and assume spherical plasmoids (blobs) as representative structures. Each plasmoid shares the same physical parameters (see Table \ref{tab:1} -- 
 electron luminosity and magnetic field strength decrease linearly with time) but follows a particle density distribution trend inspired by Fig.\ref{Fig:rho}. This trend implies that the density distribution is asymmetric relative to the disk midplane, likely due to perturbations caused by the secondary black hole passing through the accretion disk. 

The emission from each plasmoid was calculated as it evolved and expanded, and we normalized this emission by scaling to match the observed total number of plasmoids. These plasmoid families, introduced at discrete time intervals, serve to capture the overall observed variability trends. At the same time, the X-rays that are produced by the blobs do not violate the steady-state emission. For the X-ray flare, we modeled a temporal increase in particle density (and consequently $L_e$) within each plasmoid, following the density evolution shown in Fig.\ref{Fig:rho}, which indicates an 80 times increase, according to the observations. In this case, the source becomes optically thin at radio frequencies due to synchrotron self-absorption (SSA).

\begin{figure}[t]
    \centering    \includegraphics[width=1.0\linewidth]{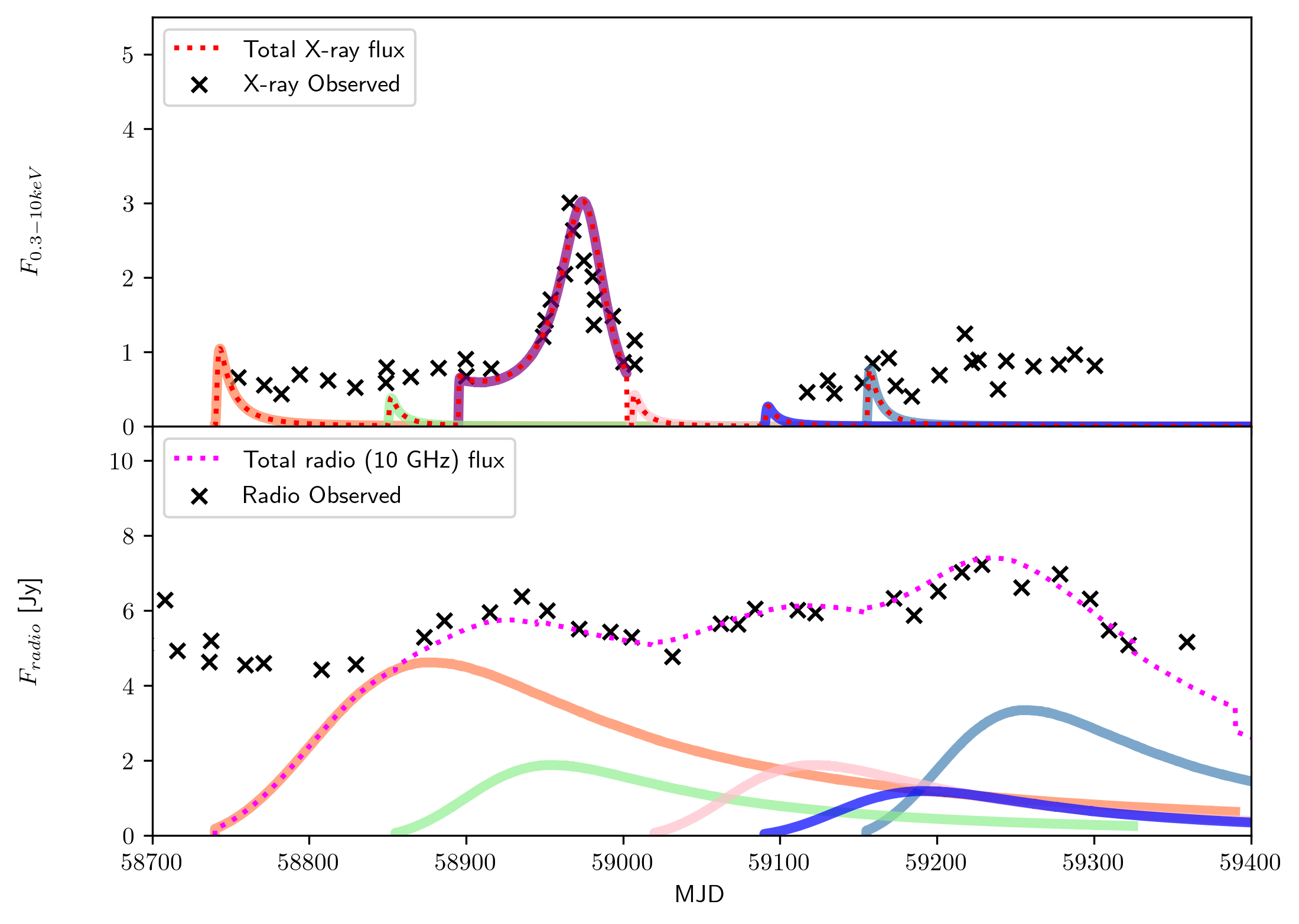}
    \caption{Radio and X-ray lightcurves of OJ287 variability in 2019, the points represent the observational trend. We show the total emission from different blob families as they evolve in time. }
    \label{Fig:modeling}
\end{figure}

\section{Results and discussion}
\label{sec:results}
To explain the variability of OJ287, several models have been proposed, 
including a varying accretion rate, a low-density cavity, and a precessing jet 
\citealt{Katz1997, Tanaka2013, Britzen2018}. While these 
models offer insights, significant attention has been given to the high-energy 
emission from blazars.
Our analysis focuses on the period from 2019 to 2021, covering a crucial phase of 
the secondary black hole's orbit. The starting point corresponds to the secondary's passage 
through the primary's accretion disk, while the concluding phase occurs as the secondary 
moves near the primary's relativistic jet. Notably, an emission event is observed around 
2020, coinciding with the time frame in which the secondary’s trajectory brings it into a 
magnetized environment influenced by both the disk and jet regions.

The formation and evolution of plasmoids are described in section \ref{sec:plasm}, while 
the emission mechanisms are detailed in section \ref{sec:emis}. 
Our results suggest that the primary emission spot is located approximately 5000 AU, 
corresponding to a distance of $10^{16.9}$ cm from the primary black hole. Additionally, we considered 
a secondary spot at 3250 AU, corresponding to $10^{16.7}$ cm.

The derived blob sizes are roughly an order of magnitude smaller than the primary black 
hole's Schwarzschild radius (almost an order of magnitude larger than the secondary). 
The radio emission likely originates from an optically thin region, where SSA becomes less significant and radio photons can escape. If this emission stems from a single 
plasmoid, we expect a region of approximately $10^{17}$ cm. However, this would violate 
the event horizon constraints of the black hole, supporting the concept of a plasmoid 
chain.

In our scenario, we estimate around a thousand plasmoids in total. Some of them are produced during phase 1, 
when the secondary black hole penetrates the primary’s accretion disk. 
This specific interaction corresponds to the equatorial crossing that occurred around 
2019\footnote{The phase angle, $\phi_i$, associated with this crossing is approximately $313^{\circ}$, derived from the quasi-Keplerian model parameters: an orbital period of 12.13 years, an eccentricity of $e=0.65$, and a pericenter precession of $38^{\circ}$ per orbit. The timing of this crossing was calculated using the relation:
$T(\phi_i) = \frac{T_{\rm orb}}{2\pi} \left( 2 \arctan\left(0.46 \tan\left(\frac{\phi_i}{2}\right)\right) 
- \frac{0.598 \tan\left(\frac{\phi_i}{2}\right)}{1 + 0.2116 \tan^2\left(\frac{\phi_i}{2}\right)} \right), $
with \(\phi_i\) linked to the forward precession per orbit. This event occurred near the epoch 2019.57. These parameters are critical for determining the precise timing and geometry of the secondary black hole’s disk crossings \citet{Valtonen2023c, Valtonen2024}.}  and is at the beginning of the observations in Fig. \ref{Fig:modeling}, which aligns with the quasi-Keplerian orbital sequence described by \citet{Valtonen2023c, Valtonen2024}.
Plasmoid properties vary, leading to different contributions to the overall variability. 
This zero-order approach provides a framework for determining the maximum plasmoid size and the minimum number required for such variability. According to the plasmoid creation rate in Eq. \ref{dotN}, the plasmoid population can grow over time, explaining the observed fluctuations. As the particle density increases, we can account for the onset of the X-ray flare. The region responsible for these flares remains optically thick to radio frequencies due to SSA.
During phase 2, as the secondary black hole exits the disk, the overall luminosity decreases, followed by another radio flare, similar to phase 1. Table \ref{tab:1} summarizes the averaged parameter values used in this analysis, which align well with the plasmoid properties described in section \ref{sec:plasm}.

\begin{figure}[t]
    \centering    \includegraphics[width=1.0\linewidth]{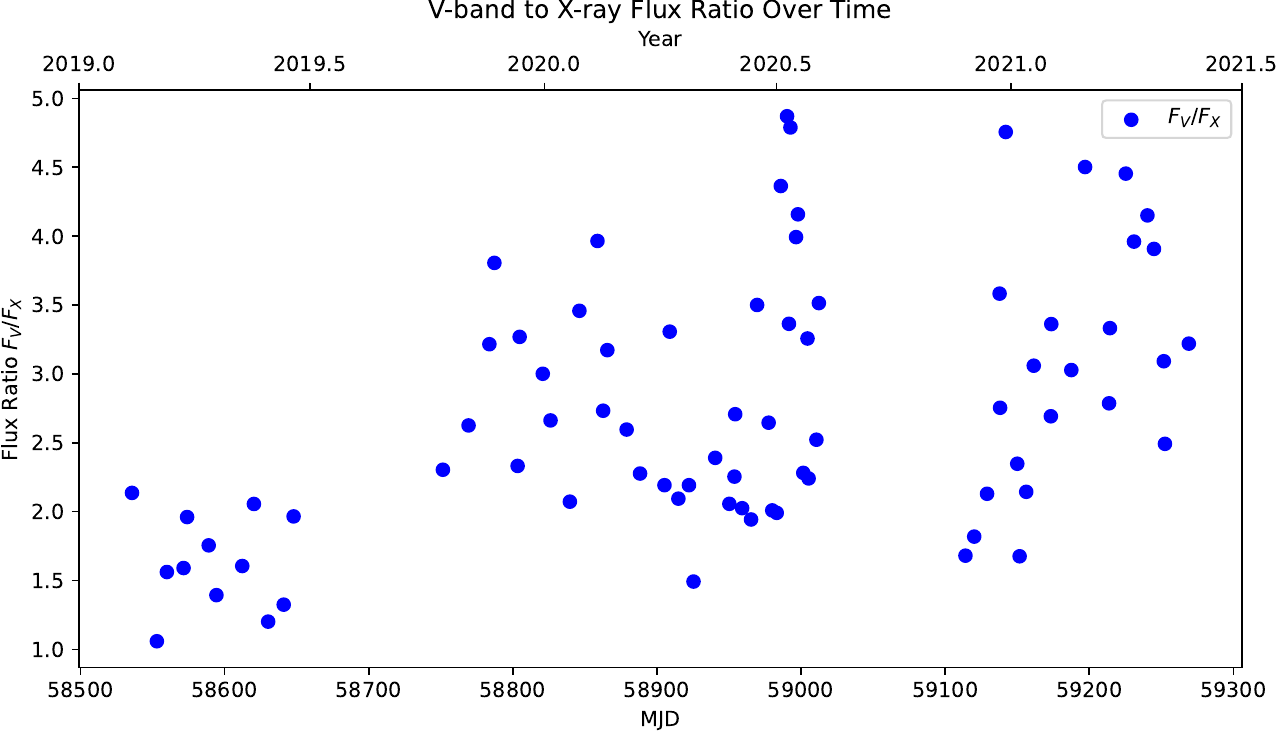}
    \caption{Ratio of optical (V-band) to X-ray flux ($F_V/F_X$) for OJ 287 as a function of time. The data points were derived from Swift observations and ground-based V-band measurements. 
    }
    \label{Fig:anomaly}
\end{figure}

Figure \ref{Fig:modeling} shows the modeled lightcurve of X-ray 
and radio emission. Our model suggests that high-energy photons escape first, followed 
by lower-frequency emission. While the radio emission from the X-ray-producing region 
is not prominent in the figure due to SSA, the overall emission trend remains  
consistent. These results are indicative, and various combinations of plasmoid 
properties can account for the total variability observed in the source.
Figure~\ref{Fig:anomaly} highlights anomalous $F_V / F_X$ values that deviate from the typical jet-dominated emission range. These anomalies coincide with the secondary black hole’s passage through the accretion disk of the primary in 2019 and its subsequent interaction near the jet region in 2020. The V-band to X-ray flux ratio during this period, plotted in Fig.~\ref{Fig:anomaly}, exhibits significant deviations following the secondary’s passage. This behavior aligns with the expectations from the V/X method \citep{Valtonen2021, Valtonen2023}, which identifies departures from the standard jet emission profile. In particular, the shift in the V/X ratio suggests that an additional emission mechanism, distinct from the steady jet emission, may be at play. One possible explanation for these deviations is the contribution of magnetic reconnection-driven plasmoid formation, which could enhance variability during these phases.

It is important to note that this is a simplified or “toy” model. We deliberately avoid complex mechanisms such as bremsstrahlung, which have been discussed in the literature {\citep{2012Valtonen,2020ApJ...898...50Y}}, and do not explore the detailed role of the accretion disk. While the disk's structure may influence our results, it is outside the scope of this study. For instance, studies like \citep{TMG1998} have 
explored the origins of high-energy radiation in blazar jets and given constraints on the physical parameters 
that are relevant to understanding OJ 287’s multiwavelength variability. Future work could further explore how disk properties impact the plasmoid dynamics and emission processes.

\section{Conclusions}
\label{sec:con}
In this letter, we present a simple model to explain the variability of OJ 287 through the formation of plasmoid chains. Beyond the observed variability, our model also provides a possible explanation for the anomalous V/X flux ratio, suggesting that magnetic reconnection during the secondary's passage contributes to additional emission components.

\begin{itemize}

    \item As the secondary black hole passes through the accretion disk and subsequently moves close to the jet of the primary, it disturbs the magnetic geometry, inducing shocks and triggering intense magnetic reconnection, leading to the formation of plasmoids and plasmoid chains.
    
    \item From the perspective of the secondary black hole, this magnetic reconnection process appears as though a large magnetic forest is being accreted, contrasting with the usual inward accretion of matter.

    \item Our numerical simulations suggest plasmoids' formation and size distribution, providing a plausible confirmation of the model, consistent with values found in the literature.

    \item A simple approach combining synchrotron emission and inverse Compton scattering can account for the observed variability in OJ 287.

    \item We acknowledge that in a more realistic scenario the size of the plasmoids is stochastic and the required energy for such plasmoids is variable. However, our model presents a general trend.

\end{itemize}

\begin{acknowledgements}
We would like to thank {the anonymous referee, for their constructive comments}. Furthermore, we thank Apostolos Mastichiadis and Fabrizio Tavecchio for the fruitful discussion and comments. This work was supported by computational time granted from the National Infrastructures for Research and Technology S.A. (GRNET S.A.) in the National HPC facility - ARIS - under project ID 16033.
\end{acknowledgements}

\bibliographystyle{aa} 
\bibliography{oj287}

\begin{thebibliography}{53}
\expandafter\ifx\csname natexlab\endcsname\relax\def\natexlab#1{#1}\fi

\bibitem[{{Ball} {et~al.}(2018{\natexlab{a}}){Ball}, {Sironi}, \&
  {{\"O}zel}}]{BAO2018}
{Ball}, D., {Sironi}, L., \& {{\"O}zel}, F. 2018{\natexlab{a}}, \apj, 862, 80

\bibitem[{{Ball} {et~al.}(2018{\natexlab{b}}){Ball}, {Sironi}, \&
  {{\"O}zel}}]{Ball2018}
{Ball}, D., {Sironi}, L., \& {{\"O}zel}, F. 2018{\natexlab{b}}, \apj, 862, 80

\bibitem[{Blandford {et~al.}(2019)Blandford, Meier, \&
  Readhead}]{Blandford2019}
Blandford, R., Meier, D., \& Readhead, A. 2019, Annual Review of Astronomy and
  Astrophysics, 57, 467

\bibitem[{{Boula} \& {Mastichiadis}(2022)}]{BM2022}
{Boula}, S. \& {Mastichiadis}, A. 2022, \aap, 657, A20

\bibitem[{{Britzen} {et~al.}(2018){Britzen}, {Fendt}, {Witzel}, {Qian},
  {Pashchenko}, {Kurtanidze}, {Zajacek}, {Martinez}, {Karas}, {Aller}, {Aller},
  {Eckart}, {Nilsson}, {Ar{\'e}valo}, {Cuadra}, {Subroweit}, \&
  {Witzel}}]{Britzen2018}
{Britzen}, S., {Fendt}, C., {Witzel}, G., {et~al.} 2018, \mnras, 478, 3199

\bibitem[{{Chatterjee} {et~al.}(2021){Chatterjee}, {Markoff}, {Neilsen},
  {Younsi}, {Witzel}, {Tchekhovskoy}, {Yoon}, {Ingram}, {van der Klis},
  {Boyce}, {Do}, {Haggard}, \& {Nowak}}]{Chatterjee2021}
{Chatterjee}, K., {Markoff}, S., {Neilsen}, J., {et~al.} 2021, Mon. Not. R.
  Astron. Soc., 507, 5281

\bibitem[{{Fishbone} \& {Moncrief}(1976)}]{Fishborne_Moncrief_1976}
{Fishbone}, L.~G. \& {Moncrief}, V. 1976, \apj, 207, 962

\bibitem[{{Giannios}(2013)}]{Giannios2013}
{Giannios}, D. 2013, Mon. Not. R. Astron. Soc., 431, 355

\bibitem[{Gold {et~al.}(2014)Gold, Paschalidis, Etienne, Shapiro, \&
  Pfeiffer}]{Gold2014}
Gold, R., Paschalidis, V., Etienne, Z.~B., Shapiro, S.~L., \& Pfeiffer, H.~P.
  2014, Physical Review D, 89, 064060

\bibitem[{{G{\'o}mez} {et~al.}(2022){G{\'o}mez}, {Traianou},
  {et~al.}}]{Gomez2022}
{G{\'o}mez}, J.~L., {Traianou}, E., {et~al.} 2022, \apj, 924, 122

\bibitem[{{Guo} {et~al.}(2014){Guo}, {Li}, {Daughton}, \& {Liu}}]{Guo2014}
{Guo}, F., {Li}, H., {Daughton}, W., \& {Liu}, Y.-H. 2014, Phys. Rev. Lett.,
  113, 155005

\bibitem[{{Guti{\'e}rrez} {et~al.}(2024){Guti{\'e}rrez}, {Combi}, {Romero}, \&
  {Campanelli}}]{Gutierrez2024}
{Guti{\'e}rrez}, E.~M., {Combi}, L., {Romero}, G.~E., \& {Campanelli}, M. 2024,
  Mon. Not. R. Astron. Soc., 532, 506

\bibitem[{Igumenshchev(2008)}]{Igumenshchev_2008}
Igumenshchev, I.~V. 2008, The Astrophysical Journal, 677, 317

\bibitem[{Igumenshchev {et~al.}(2003)Igumenshchev, Narayan, \&
  Abramowicz}]{Igumenshchev_2003}
Igumenshchev, I.~V., Narayan, R., \& Abramowicz, M.~A. 2003, The Astrophysical
  Journal, 592, 1042

\bibitem[{{Ivanov} {et~al.}(1998){Ivanov}, {Igumenshchev}, \&
  {Novikov}}]{1998Ivanov}
{Ivanov}, P.~B., {Igumenshchev}, I.~V., \& {Novikov}, I.~D. 1998, \apj, 507,
  131

\bibitem[{{Katz}(1997)}]{Katz1997}
{Katz}, J.~I. 1997, \apj, 478, 527

\bibitem[{{Komossa} {et~al.}(2021){Komossa}, {Grupe}, {Kraus}, {Gallo},
  {Gonzalez}, {Parker}, {Valtonen}, {Hollett}, {Bach}, {G{\'o}mez}, {Myserlis},
  \& {Ciprini}}]{Komossa2021}
{Komossa}, S., {Grupe}, D., {Kraus}, A., {et~al.} 2021, Universe, 7, 261

\bibitem[{Komossa {et~al.}(2023)Komossa, Grupe, Kraus, Gurwell, Haiman, Liu,
  Tchekhovskoy, Gallo, Berton, Blandford, Gómez, \& Gonzalez}]{Komossa2023}
Komossa, S., Grupe, D., Kraus, A., {et~al.} 2023, Monthly Notices of the Royal
  Astronomical Society: Letters, 522, L84

\bibitem[{{Laine} {et~al.}(2020){Laine}, {Dey}, {Valtonen}, {Gopakumar},
  {Zola}, {Komossa}, {Kidger}, {Pihajoki}, {G{\'o}mez}, {Caton}, {Ciprini},
  {Drozdz}, {Gazeas}, {Godunova}, {Haque}, {Hildebrandt}, {Hudec}, {Jermak},
  {Kong}, {Lehto}, {Liakos}, {Matsumoto}, {Mugrauer}, {Pursimo}, {Reichart},
  {Simon}, {Siwak}, \& {Sonbas}}]{Laine2020}
{Laine}, S., {Dey}, L., {Valtonen}, M., {et~al.} 2020, \apjl, 894, L1

\bibitem[{{Lehto} \& {Valtonen}(1996)}]{1996Lehto}
{Lehto}, H.~J. \& {Valtonen}, M.~J. 1996, \apj, 460, 207

\bibitem[{{Li} {et~al.}(2023){Li}, {Guo}, {Liu}, \& {Li}}]{LGLL2023}
{Li}, X., {Guo}, F., {Liu}, Y.-H., \& {Li}, H. 2023, \apjl, 954, L37

\bibitem[{{Marscher} \& {Jorstad}(2011)}]{Marscher2011}
{Marscher}, A.~P. \& {Jorstad}, S.~G. 2011, Astrophys. J., 729, 26

\bibitem[{{McKinney} \& {Gammie}(2004)}]{McKinney_2004}
{McKinney}, J.~C. \& {Gammie}, C.~F. 2004, \apj, 611, 977

\bibitem[{{Meringolo} {et~al.}(2023){Meringolo}, {Cruz-Osorio}, {Rezzolla}, \&
  {Servidio}}]{Meringolo2023}
{Meringolo}, C., {Cruz-Osorio}, A., {Rezzolla}, L., \& {Servidio}, S. 2023,
  \apj, 944, 122

\bibitem[{Mizuno {et~al.}(2014)Mizuno, Hardee, \& Nishikawa}]{Mizuno2014}
Mizuno, Y., Hardee, P.~E., \& Nishikawa, K.-I. 2014, The Astrophysical Journal,
  784, 167
\bibitem[{{Myserlis } {et~al.}(2018){Myserlis }, {Komossa, S.}, {Angelakis,
  E.}, {Gómez, J. L.}, {Karamanavis, V.}, {Krichbaum, T. P.}, {Bach, U.}, and {Grupe, D.}}]{Myserlis2018}
{Myserlis }, {Komossa, S.}, {Angelakis, E.}, {et~al.} 2018, \aap, 619, A88


\bibitem[{{Nathanail} {et~al.}(2020){Nathanail}, {Fromm}, {Porth}, {Olivares},
  {Younsi}, {Mizuno}, \& {Rezzolla}}]{2020NFPOYMR}
{Nathanail}, A., {Fromm}, C.~M., {Porth}, O., {et~al.} 2020, \mnras, 495, 1549

\bibitem[{{Nathanail} {et~al.}(2022){Nathanail}, {Mpisketzis}, {Porth},
  {Fromm}, \& {Rezzolla}}]{2022NMPFR}
{Nathanail}, A., {Mpisketzis}, V., {Porth}, O., {Fromm}, C.~M., \& {Rezzolla},
  L. 2022, \mnras, 513, 4267

\bibitem[{{Olivares} {et~al.}(2019){Olivares}, {Porth}, {Davelaar}, {Most},
  {Fromm}, {Mizuno}, {Younsi}, \& {Rezzolla}}]{Olivares_2019}
{Olivares}, H., {Porth}, O., {Davelaar}, J., {et~al.} 2019, \aap, 629, A61

\bibitem[{{Petropoulou} {et~al.}(2016){Petropoulou}, {Giannios}, \&
  {Sironi}}]{Petropoulou2016}
{Petropoulou}, M., {Giannios}, D., \& {Sironi}, L. 2016, Mon. Not. R. Astron.
  Soc., 462, 3325

\bibitem[{Pihajoki {et~al.}(2013)Pihajoki, Valtonen, \& Ciprini}]{Pihajoki2013}
Pihajoki, P., Valtonen, M., \& Ciprini, S. 2013, Monthly Notices of the Royal
  Astronomical Society, 434, 3122

\bibitem[{{Porth} {et~al.}(2017){Porth}, {Olivares}, {Mizuno}, {Younsi},
  {Rezzolla}, {Moscibrodzka}, {Falcke}, \& {Kramer}}]{Porth_2017}
{Porth}, O., {Olivares}, H., {Mizuno}, Y., {et~al.} 2017, Computational
  Astrophysics and Cosmology, 4, 1

\bibitem[{{Ripperda} {et~al.}(2020){Ripperda}, {Bacchini}, \&
  {Philippov}}]{Ripperda2020}
{Ripperda}, B., {Bacchini}, F., \& {Philippov}, A.~A. 2020, Astrophys. J., 900,
  100

\bibitem[{{Ripperda} {et~al.}(2022){Ripperda}, {Liska}, {Chatterjee}, {Musoke},
  {Philippov}, {Markoff}, {Tchekhovskoy}, \& {Younsi}}]{Ripperda2022}
{Ripperda}, B., {Liska}, M., {Chatterjee}, K., {et~al.} 2022, Astrophys. J.
  Lett., 924, L32

\bibitem[{{Rybicki} \& {Lightman}(1979)}]{RL1979}
{Rybicki}, G.~B. \& {Lightman}, A.~P. 1979, {Radiative processes in
  astrophysics}

\bibitem[{{Sillanpaa} {et~al.}(1988){Sillanpaa}, {Haarala}, {Valtonen},
  {Sundelius}, \& {Byrd}}]{Sillanpaa1988}
{Sillanpaa}, A., {Haarala}, S., {Valtonen}, M.~J., {Sundelius}, B., \& {Byrd},
  G.~G. 1988, \apj, 325, 628

\bibitem[{{Sironi} {et~al.}(2016){Sironi}, {Giannios}, \&
  {Petropoulou}}]{Sironi2016}
{Sironi}, L., {Giannios}, D., \& {Petropoulou}, M. 2016, Mon. Not. R. Astron.
  Soc., 462, 48

\bibitem[{{Sironi} \& {Spitkovsky}(2014)}]{Sironi2014}
{Sironi}, L. \& {Spitkovsky}, A. 2014, Astrophys. J.l, 783, L21

\bibitem[{{Tanaka}(2013)}]{Tanaka2013}
{Tanaka}, T.~L. 2013, \mnras, 434, 2275

\bibitem[{{Tavecchio} {et~al.}(1998){Tavecchio}, {Maraschi}, \&
  {Ghisellini}}]{TMG1998}
{Tavecchio}, F., {Maraschi}, L., \& {Ghisellini}, G. 1998, \apj, 509, 608

\bibitem[{Tchekhovskoy {et~al.}(2011)Tchekhovskoy, Narayan, \&
  McKinney}]{Tchekhovskoy_2011}
Tchekhovskoy, A., Narayan, R., \& McKinney, J.~C. 2011, Monthly Notices of the
  Royal Astronomical Society: Letters, 418, L79

\bibitem[{{Valtonen} {et~al.}(2012){Valtonen}, {Ciprini}, \&
  {Lehto}}]{2012Valtonen}
{Valtonen}, M.~J., {Ciprini}, S., \& {Lehto}, H.~J. 2012, \mnras, 427, 77

\bibitem[{{Valtonen} {et~al.}(2021){Valtonen}, {Dey}, {Gopakumar}, {Zola},
  {Komossa}, {Pursimo}, {Gomez}, {Hudec}, {Jermak}, \&
  {Berdyugin}}]{Valtonen2021}
{Valtonen}, M.~J., {Dey}, L., {Gopakumar}, A., {et~al.} 2021, Galaxies, 10, 1

\bibitem[{{Valtonen} {et~al.}(2023{\natexlab{a}}){Valtonen}, {Dey},
  {Gopakumar}, {Zola}, {L{\"a}hteenm{\"a}ki}, {Tornikoski}, {Gupta}, {Pursimo},
  {Knudstrup}, {Gomez}, {Hudec}, {Jel{\'\i}nek}, {{\v{S}}trobl}, {Berdyugin},
  {Ciprini}, {Reichart}, {Kouprianov}, {Matsumoto}, {Drozdz}, {Mugrauer},
  {Sadun}, {Zejmo}, {Sillanp{\"a}{\"a}}, {Lehto}, {Nilsson}, {Imazawa}, \&
  {Uemura}}]{Valtonen2023}
{Valtonen}, M.~J., {Dey}, L., {Gopakumar}, A., {et~al.} 2023{\natexlab{a}},
  Galaxies, 11, 82

\bibitem[{Valtonen {et~al.}(2008)Valtonen, Lehto, Nilsson, Heidt, Takalo,
  Sillanp{\"a}{\"a}, Villforth, Kidger, Poyner, Pursimo,
  {et~al.}}]{Valtonen2008}
Valtonen, M.~J., Lehto, H., Nilsson, K., {et~al.} 2008, Nature, 452, 851

\bibitem[{{Valtonen} {et~al.}(2016){Valtonen}, {Zola}, {Ciprini}, {Gopakumar},
  {Matsumoto}, {Sadakane}, {Kidger}, {Gazeas}, {Nilsson}, {Berdyugin},
  {Piirola}, {Jermak}, {Baliyan}, {Alicavus}, {Boyd}, {Campas Torrent},
  {Campos}, {Carrillo G{\'o}mez}, {Caton}, {Chavushyan}, {Dalessio}, {Debski},
  {Dimitrov}, {Drozdz}, {Er}, {Erdem}, {Escartin P{\'e}rez}, {Fallah Ramazani},
  {Filippenko}, {Ganesh}, {Garcia}, {G{\'o}mez Pinilla}, {Gopinathan},
  {Haislip}, {Hudec}, {Hurst}, {Ivarsen}, {Jelinek}, {Joshi}, {Kagitani},
  {Kaur}, {Keel}, {LaCluyze}, {Lee}, {Lindfors}, {Lozano de Haro}, {Moore},
  {Mugrauer}, {Naves Nogues}, {Neely}, {Nelson}, {Ogloza}, {Okano}, {Pandey},
  {Perri}, {Pihajoki}, {Poyner}, {Provencal}, {Pursimo}, {Raj}, {Reichart},
  {Reinthal}, {Sadegi}, {Sakanoi}, {Salto Gonz{\'a}lez}, {Sameer}, {Schweyer},
  {Siwak}, {Sold{\'a}n Alfaro}, {Sonbas}, {Steele}, {Stocke}, {Strobl},
  {Takalo}, {Tomov}, {Tremosa Espasa}, {Valdes}, {Valero P{\'e}rez},
  {Verrecchia}, {Webb}, {Yoneda}, {Zejmo}, {Zheng}, {Telting}, {Saario},
  {Reynolds}, {Kvammen}, {Gafton}, {Karjalainen}, {Harmanen}, \&
  {Blay}}]{2016Valtonen}
{Valtonen}, M.~J., {Zola}, S., {Ciprini}, S., {et~al.} 2016, \apjl, 819, L37

\bibitem[{{Valtonen} {et~al.}(2023{\natexlab{b}}){Valtonen}, {Zola},
  {Gopakumar}, {L{\"a}hteenm{\"a}ki}, {Tornikoski}, {Dey}, {Gupta}, {Pursimo},
  {Knudstrup}, {Gomez}, {Hudec}, {Jel{\'\i}nek}, {{\v{S}}trobl}, {Berdyugin},
  {Ciprini}, {Reichart}, {Kouprianov}, {Matsumoto}, {Drozdz}, {Mugrauer},
  {Sadun}, {Zejmo}, {Sillanp{\"a}{\"a}}, {Lehto}, {Nilsson}, {Imazawa}, \&
  {Uemura}}]{Valtonen2023c}
{Valtonen}, M.~J., {Zola}, S., {Gopakumar}, A., {et~al.} 2023{\natexlab{b}},
  \mnras, 521, 6143

\bibitem[{{Valtonen} {et~al.}(2023{\natexlab{c}}){Valtonen}, {Zola},
  {Gopakumar}, {L{\"a}hteenm{\"a}ki}, {Tornikoski}, {Dey}, {Gupta}, {Pursimo},
  {Knudstrup}, {Gomez}, {Hudec}, {Jel{\'\i}nek}, {{\v{S}}trobl}, {Berdyugin},
  {Ciprini}, {Reichart}, {Kouprianov}, {Matsumoto}, {Drozdz}, {Mugrauer},
  {Sadun}, {Zejmo}, {Sillanp{\"a}{\"a}}, {Lehto}, {Nilsson}, {Imazawa},
  {Uemura}, \& {Davidson}}]{2023Valtonenb}
{Valtonen}, M.~J., {Zola}, S., {Gopakumar}, A., {et~al.} 2023{\natexlab{c}},
  \mnras, 525, 1153

\bibitem[{{Valtonen} {et~al.}(2024){Valtonen}, {Zola}, {Gupta}, {Kishore},
  {Gopakumar}, {Jorstad}, {Wiita}, {Gu}, {Nilsson}, {Marscher}, {Zhang},
  {Hudec}, {Matsumoto}, {Drozdz}, {Ogloza}, {Berdyugin}, {Reichart},
  {Mugrauer}, {Dey}, {Pursimo}, {Lehto}, {Ciprini}, {Nakaoka}, {Uemura},
  {Imazawa}, {Zejmo}, {Kouprianov}, {Davidson}, {Sadun}, {{\v{S}}trobl},
  {Weaver}, \& {Jel{\'\i}nek}}]{Valtonen2024}
{Valtonen}, M.~J., {Zola}, S., {Gupta}, A.~C., {et~al.} 2024, \apjl, 968, L17

\bibitem[{{Valtonen} {et~al.}(2019){Valtonen}, {Zola}, {Pihajoki}, {Enestam},
  {et~al.}}]{Valtonen2019}
{Valtonen}, M.~J., {Zola}, S., {Pihajoki}, P., {Enestam}, S., {et~al.} 2019,
  Astrophys. J., 882, 88

\bibitem[{{Werner} {et~al.}(2018){Werner}, {Uzdensky}, {Begelman}, {Cerutti},
  \& {Nalewajko}}]{Werner2018}
{Werner}, G.~R., {Uzdensky}, D.~A., {Begelman}, M.~C., {Cerutti}, B., \&
  {Nalewajko}, K. 2018, Mon. Not. R. Astron. Soc., 473, 4840

\bibitem[{{Yarza} {et~al.}(2020){Yarza}, {Wong}, {Ryan}, \&
  {Gammie}}]{2020ApJ...898...50Y}
{Yarza}, R., {Wong}, G.~N., {Ryan}, B.~R., \& {Gammie}, C.~F. 2020, \apj, 898,
  50

\bibitem[{{Zhang} {et~al.}(2024){Zhang}, {Dong}, \& {Giannios}}]{Zhang2024}
{Zhang}, H., {Dong}, L., \& {Giannios}, D. 2024, Mon. Not. R. Astron. Soc.,
  531, 4781

\end{thebibliography}

\newpage
\begin{appendix}
\section{Numerical simulation setup}
\label{appA}
In this study, the simulations are conducted in 3D space using the 
BHAC code (\cite{Porth_2017,Olivares_2019})\footnote{, https://www.bhac.science}, 
which employs modified Kerr-Schild coordinates (\cite{McKinney_2004}) and 
incorporates 2–3 levels of adaptive mesh refinement. 
The simulation units are chosen 
such that ${\rm c = G }= 1$, meaning the black hole mass is used as the unit of length, 
with the gravitational radius $r_{\rm g} = M$, and the grid resolution is set to $384 
\times 192 \times 192$ cells along the $r$, $\theta$, and $\phi$ axes, 
respectively.

The initial setup consists of a torus in hydrodynamic equilibrium, as described by 
\cite{Fishborne_Moncrief_1976}, with a constant specific angular momentum of $l = 
6.76$, orbiting a Kerr black hole with a dimensionless spin parameter $a = 0.94$. 
The disk's inner edge is located at $r_{\text{in}} = 20 M$, while the maximum 
density is at $r_{\text{max}} = 40 M$. The equation of state is ideal with an 
adiabatic index of $\gamma = 4/3$. Additionally, the magnetic field is initialized 
as a nested loop structure, defined by the vector potential:

\begin{equation}
\label{vector_potential}
    A_{\phi} = max \left( \left( \dfrac{\rho}{\rho_{max}} \, \left(\dfrac{r}{r_{in}}\right)^3 \, \sin^3\theta\,\exp{\left(\dfrac{-r}{400}\right)} \right) - 0.2, 0 \right),
\end{equation}
where $\rho_{max}$ is the maximum rest-mass density in the torus.

The simulations evolve into a Magnetically Arrested Disk (MAD) state, where the 
poloidal magnetic flux is advected toward the black hole by the accreting gas. As 
the magnetic pressure builds up, it eventually becomes strong enough to 
counterbalance the ram pressure of the accreting disk material 
\citep{Igumenshchev_2003, Igumenshchev_2008, Tchekhovskoy_2011}. At this point, 
when the magnetic pressure and ram pressure reach an equipartition state, gas 
accretion is significantly hindered, and further accretion can only occur via 
3D non-axisymmetric processes. 
To ensure accurate density distributions, the simulation is allowed to run until 
it reaches a quasi-stationary MAD state. By time $30000M$, we are confident that 
the system has reached this state, and at this point, the data is exported for 
analysis. Specifically, the exported data is used to generate Fig. \ref{Fig:mhd}, 
which provides a 3D visualization of the accretion disk, and Fig. \ref{Fig:rho}, which shows the vertical density distribution for different radial distances across the disk.

\section{Details for plasmoid chains}
\label{appB}

Plasmoids that evolve within highly magnetized environments serve as efficient sites for 
particle acceleration, producing high-energy emissions consistent with the observed 
variability in blazar jets \citep{Sironi2016}. This process aligns well with our model for 
OJ 287, where the passage of the secondary black hole through the primary’s magnetized 
accretion disk disturbs the magnetic field, triggering reconnection events and the 
formation of plasmoid chains. The particle acceleration within these plasmoids can explain 
the nonthermal X-ray and radio flares observed in the system. Additionally, plasmoid 
mergers during magnetic reconnection can produce significant multiwavelength flares and 
polarization variations \citep{Zhang2024}.

Plasmoids are frequently produced in reconnection layers during magnetic reconnection 
events in turbulent black hole accretion flows.  
Recent simulations indicate that as 
magnetic field lines are advected toward the black hole, plasmoids frequently form at the 
boundaries of highly magnetized funnel regions \citep{2020NFPOYMR, Ripperda2020, 
Ripperda2022, 2022NMPFR}. Their formation is closely tied to current sheet structures that 
develop and fragment during the accretion process near the black hole, typically in a 
region spanning $2 - 15 ~ r_{\rm g}$. These simulations show that magnetic reconnection near the 
black hole can efficiently produce macroscopic, filamentary plasmoids with lengths of $
\sim 10 - 15 ~ r_{\rm g}$ and diameters of $\sim 3 - 5 ~ r_{\rm g}$. They are generated approximately 
every $500 - 700~  t_{\rm g}$ and generally do not survive as compact structures beyond a 
spherical region of $\approx 30 ~ r_{\rm g}$ from the event horizon. Smaller plasmoids, with 
lengths of $4 - 6 ~ r_{\rm g}$ and diameters of $\sim 2 - 3 ~ r_{\rm g}$, are produced more 
frequently, approximately every $10 - 20 ~ t_{\rm g}$ \citep{2022NMPFR}.
As plasmoids grow, they accumulate internal energy, expanding due to the decreasing 
external pressure and density, which accelerates them. The most energetic plasmoids 
eventually become unbound from the black hole's gravitational influence. Large plasmoids 
can reach radii of up to $\approx 50 ~ r_{\rm g}$ during their evolution.

The frequency and energy budget of plasmoid production are critical to explaining the high-
energy variability observed in OJ 287. GRMHD simulations and theoretical models suggest 
that the plasmoid formation rate scales with the black hole's gravitational timescale. Over a period of $10^5$ to $2\times 10^5$ gravitational timescales ($t_{g2}$), 
approximately 1000 plasmoids are produced. Both simulations support this plasmoid production rate and prior studies on magnetic reconnection in relativistic environments 
\citep{Giannios2013, Petropoulou2016}. 
Equation \ref{dotN} suggests that a new plasmoid forms approximately every 500 $t_{\rm g2}$ which would 
correspond to 5 $t_{\rm g1}$. Note that this estimate refers to much larger plasmoids 
comparable in size to the secondary black hole. Smaller structures form more frequently 
but have negligible energy and are not considered here.

\section{Emission modelling}\label{appC}
Using the code from \cite{BM2022}, which solves the kinetic electron kinetic equations under the assumptions presented in the main text of this paper and the specified environmental conditions, we construct a self-consistent one-zone expanding leptonic model to study the connection between radio emission and higher frequency emission. The kinetic equation of electrons reads:
\begin{center} 
\begin{equation}
\frac{\partial n({\gamma},t)}{\partial t}+\frac{\partial \left[{L(\gamma,t)}n(\gamma,t)\right]}{\partial {\gamma}}+\frac{n(\gamma,t)}{t_{\rm esc}} =Q_e({\gamma},~t),
\end{equation}
\end{center}
where $n({\gamma},t)$ the differential number density of electrons of energy $\gamma$ in the interval $d\gamma$, and ${L(\gamma,t)}$ represents the loss term, which include synchrotron losses $A_{\rm syn}$, inverse Compton scattering losses $A_{\rm ICS}$, 
and adiabatic expansion losses ~$A_{\rm exp}$ and reads:
${L(\gamma,t)}=(A_{\rm syn}({\gamma},~t)+A_{\rm ICS}({\gamma},~t)+A_{\rm exp}({\gamma},~t))$.
 The magnetic field strength is derived from the source's accretion power, assuming equipartition with the kinetic energy density of the accreted mass.

The luminosity of injected electrons: 
\begin{equation}
 L_{\rm inj}^e(t)=m_{\rm e}c^2\int_{\gamma_{\rm min}}^{\gamma_{\rm max}} Q_{\rm e}(\gamma,t) \gamma {\rm{d}}\gamma=\eta_{\rm e} P_{\rm acc}, 
\end{equation}
where $\eta_{\rm e}$ is a proportionality constant, and $\gamma_{\rm min}$, $\gamma_{\rm max}$ are the minimum and maximum electron Lorentz factors.
\end{appendix}

\end{document}